\documentclass{mem}
\usepackage{natbib}\usepackage{txfonts}\usepackage{balance}
\usepackage{graphicx}
\usepackage[a4paper,breaklinks,dvipdfm]{hyperref}
\idline{87}{1}
\begin{document}
\def\teff{$T\rm_{eff }$}
\def\kms{$\mathrm {km s}^{-1}$}

\title{
Recycled Pulsars: Spins, Masses and Ages
}

   \subtitle{}

\author{T.M. \,Tauris\inst{1,2} 
        }

\institute{
Max-Planck-Institut f\"ur Radioastronomie, Auf dem H\"ugel 69, 53121 Bonn, Germany
\and
Argelander-Institut f\"ur Astronomie, Universit\"at Bonn, Auf dem H\"ugel 71, 53121 Bonn, Germany
\email{tauris@astro.uni-bonn.de}
}

\authorrunning{T.M. Tauris}

\titlerunning{Recycled Pulsars}

\abstract{
Recycled pulsars are mainly characterized by their spin periods, B-fields and masses. All these quantities are affected by previous interactions with 
a companion star in a binary system. Therefore, we can use these quantities as fossil records and learn about binary evolution. 
Here, I briefly review the distribution of these observed quantities and summarize our current understanding of the pulsar recycling process. 

}
\maketitle{}

\section{Introduction}
Recycled pulsars, or millisecond pulsars (MSPs), represent the advanced phase of stellar evolution in close, interacting binaries. Their observed
orbital and stellar properties are fossil records of their evolutionary history and thus one can use these systems as key probes of
stellar astrophysics \citep{bv91,tau11,tlk12}. 
The recycled pulsar is an old neutron star and the first formed compact object in the present-day observed binary system.
This neutron star was spun up to a high spin frequency via accretion of mass and angular momentum once the secondary star evolved. 
In this recycling phase the system is observable as a low-mass X-ray binary \citep[LMXB, e.g.][]{bv91,pw12}.
Over the last four decades, the number of known recycled pulsars has increased to $>300$ \citep{mhth05}, of which $\sim\!200$ are in binaries. 
The remaining $\sim\!100$ isolated recycled pulsars have evaporated their companion \citep{fst88,pod91}, lost it in a supernova explosion \citep{tt98} or
in an exchange encounter in a globular cluster \citep{rhs+05,frb+08}.

\section{Spins of recycled pulsars}
The observed spin periods of recycled pulsars span between 1.4~ms \citep{hrs+06} and about 200~ms \citep[e.g.][]{srm+15}.
There is a clear correlation between these spin periods and the nature of the companion star responsible
for the recycling process \citep{tlk12}. The more massive and evolved the companion star is at the onset of the mass transfer, 
the slower is the final spin rate of the recycled pulsar. The reason is that the duration of the mass-transfer (X-ray) phase is much shorter
for massive and/or giant stars, which therefore leads to less efficient recycling. 
\citet{tlk12} also derived a correlation between the final equilibrium spin period of a recycled pulsar
and the (minimum) amount of accreted material needed for spin-up.
Consider a pulsar with a typical mass of $1.4\;M_{\odot}$ and a recycled spin period of either
2, 5, 10 or 50~ms. To obtain such spins requires accretion of (at least) 0.10, 0.03, 0.01 or $10^{-3}\;M_{\odot}$, 
respectively. Therefore, it is no surprise that observed recycled pulsars with massive
companions (CO/ONeMg white dwarfs or neutron stars) are much more slow rotators, in general, compared
to MSPs with He~white dwarf companions
\begin{figure}[t!]
\resizebox{\hsize}{!}{\includegraphics[clip=true]{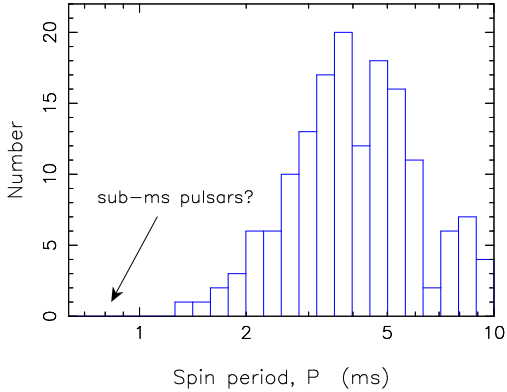}}
\caption{\footnotesize
Observed spin period distribution of MSPs. Where are the sub-ms MSPs? \citep[][]{tkb+15}.
}
\label{fig:MSP_spins}
\end{figure}

\begin{figure}[t!]
\vspace{-1.0cm}
\resizebox{\hsize}{!}{\includegraphics[clip=true]{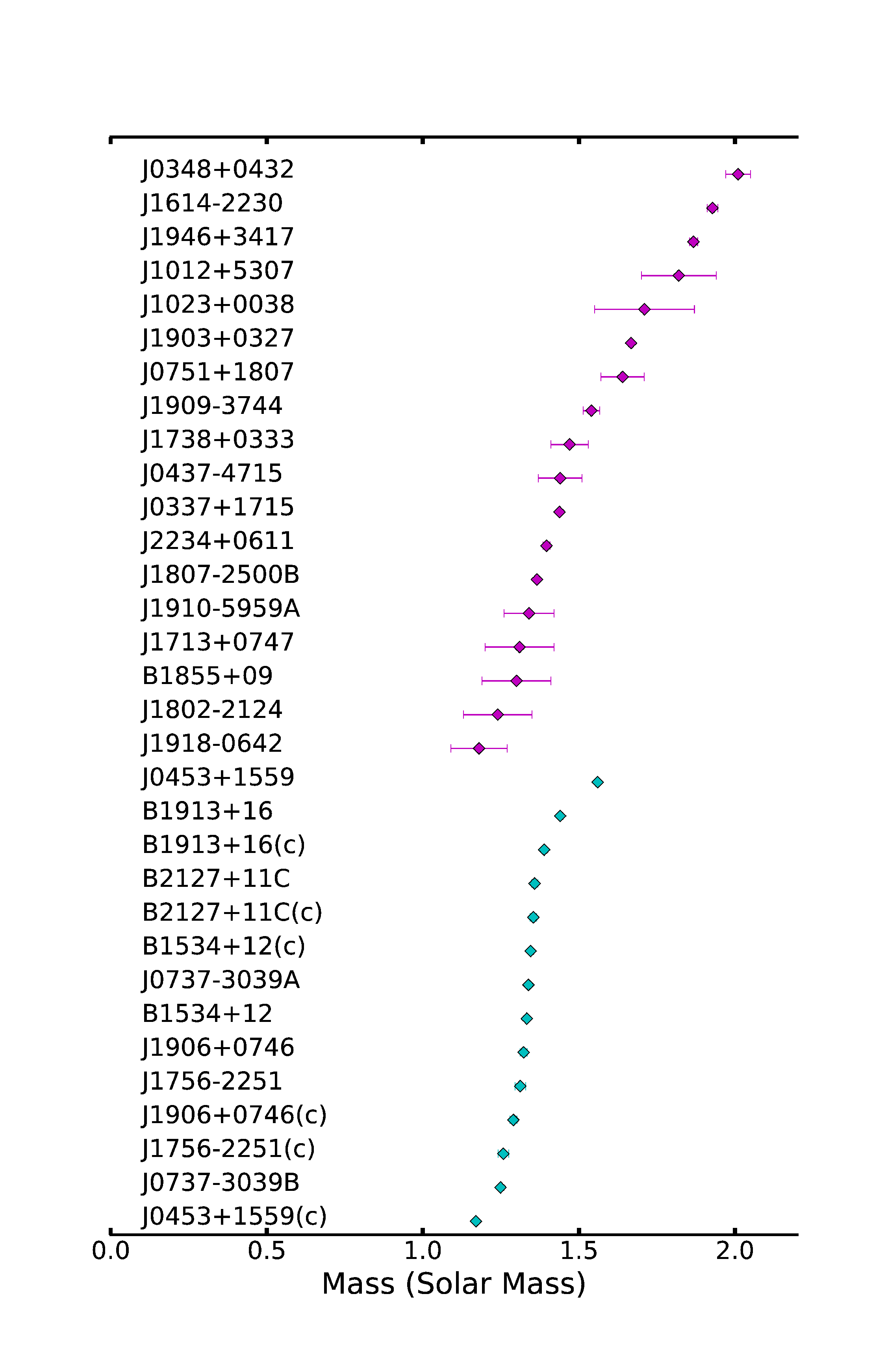}}
\caption{
\footnotesize
Mass measurements and 68\% uncertainty intervals
for binary pulsars with white dwarf companions (purple, top) and neutron star companions (blue, bottom). From \citet{atf+16}. 
}
\label{fig:MSP_masses}
\end{figure}

An interesting question is why we don't detect any MSPs spinning faster than 1.4~ms (cf. Fig.~\ref{fig:MSP_spins}). 
Answering this question is not only important for understanding magnetosphere and accretion physics, but could also be relevant for
constraining the equation-of-state of neutron stars \citep{of16}. 
Previously, the missing sub-ms MSPs were suspected to be related to Doppler smearing of radio pulsations in tight binary orbits, 
which could cause a selection bias against detection of sub-ms MSPs. However, this issue is less serious in present-day acceleration searches 
(at least for dispersion measures, $DM<100$). Three other ideas have been proposed to explain the saturation of MSP spin rates: 
i) emission of gravitational waves \citep{bil98,cmm+03}, 
ii) limited accretion torques in the pulsar magnetosphere \citep{ly05}, and
iii) spin-down at Roche lobe decoupling \citep{tau12}. 
Comparison of spin rates of accreting MSPs and radio MSPs \citep{ptrt14} gives some support for the latter hypothesis,
although no firm conclusions can be drawn. It is possible that more than one effect is at work. 

Finally, the concept of a {\it spin-up line} in the $P\dot{P}$--diagram cannot be uniquely defined, and instead one should
consider a broad {\it spin-up valley} for the birth location of recycled pulsars \citep{tlk12}. 

\section{Masses of recycled pulsars}
\begin{figure*}[t!]
\resizebox{\hsize}{!}{\includegraphics[clip=true, angle=-90]{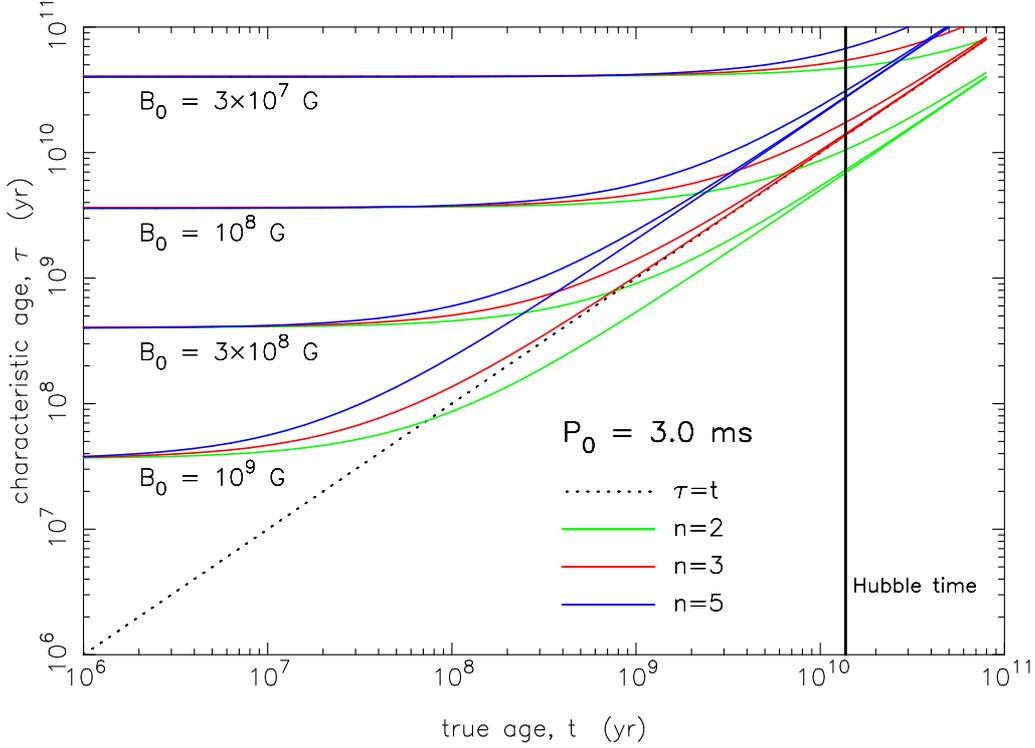}}
\caption{\footnotesize
Evolutionary tracks of characteristic ages, $\tau$, calculated as a function of true ages, $t$, for recycled pulsars evolving with a constant braking index of $n=2$, 3 and 5.
In all cases, we assumed constant values of mass, $M=1.4\;M_{\odot}$, magnetic inclination angle, $\sin\alpha = 1$, moment of inertia, $I$, 
and for $n=5$ we applied a constant ellipticity, $\epsilon\ne 0$. We assumed an initial spin period of $P_0=3.0\;{\rm  ms}$ and varied the value of the 
initial surface magnetic flux density, $B_0$. The dotted line shows a graph for $\tau = t$ and thus only pulsars located on (near) this line have characteristic ages 
as reliable age indicators. Recycled pulsars with small values of $\dot{P}_0$, resulting from small values of $B_0$, tend to have $\tau \gg t$, 
even largely exceeding the age of the Universe. See legend for identification of the various curves. From \citet{tlk12}.
}
\label{fig:MSP_ages}
\end{figure*}
In Fig.~\ref{fig:MSP_masses}, we have plotted the distribution of rather precisely measured masses
of pulsars with white dwarf and neutron star companions. 
These mass measurements span between $1.17\pm 0.01\;M_{\odot}$ \citep{msf+15} and $2.01\pm 0.04\;M_{\odot}$ \citep{afw+13}.
The largest spread in pulsar masses is seen in systems with white dwarf companions.
For the combined sample of masses, \citet{atf+16} argue in favor of a bimodal distribution with a low- and a high-mass mass peak centered at $\sim\!1.39\;M_{\odot}$ 
and $1.81\;M_{\odot}$, respectively. For binary pulsars with He white dwarf companions (the descendants of LMXBs),
one can apply the white dwarf mass--orbital period relation \citep[e.g.][]{ts99} to help constrain the mass of the neutron star.

In recent years, evidence has accumulated that accreting neutron stars are very inefficient accretors \citep[e.g.][]{avk+12,atf+16}.
Therefore, we are inclined to believe that the observed mass distribution is closely resembling the
birth distribution of neutron stars in these two different classes of pulsar binaries.
In particular, for pulsars with neutron star companions, the typical amount of accreted mass obtained from
numerical modelling \citep{tlp15} of Case~BB Roche-lobe overflow is only $\sim\!10^{-3}\;M_{\odot}$.

\section{Ages of recycled pulsars}
To understand the formation and evolution of recycled pulsars, it is of uttermost importance to determine
their true ages. Only when trues ages of MSPs are established, is it possible to gain knowledge
of pulsar evolution from the observed distribution of MSPs in the $P\dot{P}$--diagram. 
The characteristic age, $\tau \equiv P/(2\dot{P})$ is in many cases a bad estimate of true age, cf. Fig.~\ref{fig:MSP_ages}.
This is particularly the case for recycled MSPs with small values of the magnetic field. 
Since the locations of these pulsars in the $P\dot{P}$--diagram are basically {\it frozen} on a Hubble time,
they could in principle have been recycled just a few years ago -- yet they have characteristic ages
approaching 100~Gyr.   

The only way to more accurately estimate the age of a recycled pulsar is by determining the cooling age of its white dwarf companion \citep[e.g.][and references therein]{vbjj05,itla14}.

\section{Conclusions}
In this summary, I only very briefly touched on the spins, masses and ages of recycled pulsars.
New exciting pulsar discoveries at an ever-increasing rate, e.g. triple MSPs \citep{rsa+14,tv14}, eccentric MSPs \citep{crl+08,fbw+11,dsm+13}, and transitional MSPs \citep{asr+09,pfb+13} 
keep driving this field forward with fruitful results and interesting lessons to be learnt on close binary evolution
for the next many years to come. And with an almost certain guarantee for more surprises.
 
\begin{acknowledgements}
I thank the {\it Cosmic-Lab} organizer, Francesco R. Ferraro, for the invitation to the MODEST~16 conference.
I also thank the International Space Science Institute (ISSI) in Bern, Switzerland, for
funding and hosting an international team (ID~319, led by Alessandro Papitto) studying 
{\it transitional millisecond pulsars}.
\end{acknowledgements}

\bibliographystyle{aa}

\end{document}